\documentclass[twocolumn,showpacs,preprintnumbers,amsmath,amssymb]{revtex4}

\usepackage{graphicx}
\usepackage{dcolumn}
\usepackage{bm}
\usepackage{times}

\begin{document}

\preprint{}

\title{Generation of Multiple Dirac Cones in Graphene under Double-periodic and Quasiperiodic Potentials}

\author{Masayuki  Tashima}
  \email{tashima@iis.u-tokyo.ac.jp}
\affiliation{%
Department of Physics, University of Tokyo, 4-6-1 Komaba, Meguro, Tokyo 153-8505, Japan
}%

\author{Naomichi Hatano}
 \email{hatano@iis.u-tokyo.ac.jp}
\affiliation{
Institute of Industrial Science, University of Tokyo, 4-6-1 Komaba, Meguro, Tokyo 153-8505, Japan
}%

\date{\today}

\begin{abstract}
We investigate generation of new Dirac cones in graphene under double-periodic and quasiperiodic superlattice potentials.
We first show that double-periodic potentials generate the Dirac cones sporadically, following the Diophantine equation, in spite of the fact that double-periodic potentials are also periodic ones, for which previous studies predict consecutive appearance of the cones.
The sporadic appearance is due to the fact that the dispersion relation of graphene is linear only up to an energy cutoff.
We then show that quasiperiodic potentials generate the new Dirac cones densely with its density depending on the energy.
We also extend the above predictions to other materials of Dirac electrons with different energy cutoffs of  the linear dispersion.
\end{abstract}

\maketitle

\textit{Introduction}: Graphene is a monoatomic layer of carbon atoms on a hexagonal lattice \cite{review09}.
Since the ground-breaking work by Novoselov \textit{et al.} \cite{novoselov04}, graphene has attracted great interest both from theoretical and experimental points of views \cite{novoselov05, meyer07, zhang05, koshino06, fujimoto11}.
One of the remarkable features of graphene is the appearance of the massless Dirac electron.
The energy spectrum around the Fermi energy is well approximated by the Dirac Hamiltonian, which yields a linear dispersion relation, namely the Dirac cone \cite{review09}.
This has stimulated findings of Dirac cones in a wide variety of materials including $\alpha \textrm{-}( \textrm{BEDT-TTF} )_2 \textrm{I}_3$ \cite{katayama06} and $\textrm{Bi}_{0.9} \textrm{Sb}_{0.1}$ \cite{hsieh08}.

Graphene is not only studied from fundamental viewpoints, but also is expected to be a basic material for manufacturing micro-structures \cite{sun11, barbier10, zhao11}; for instance, the energy spectrum and the group velocity can be manipulated with an external superlattice potential.
The graphene superlattice has been fabricated by growing graphene on a metal surface \cite{pletikosic09} or etching patterns on graphene membranes with electron beams \cite{meyer08}.

Recent theoretical studies have revealed that graphene under a periodic external potential develops new Dirac cones in the energy spectrum around the original Dirac cone on the Fermi energy \cite{park08nat,park08}.
The new Dirac cones are predicted to appear in the dispersion at a constant interval of the reciprocal vector of the periodic potential.
The generation of the new Dirac cones are also expected to play a significant role in understanding the electric features of graphene superlattices.

In the present study, we investigate the energy spectrum of graphene under double-periodic superlattice potentials and, as a limiting case, quasiperiodic potentials.
We find for double-periodic potentials a new rule of the appearance of the new Dirac cones on the basis of the Diophantine equation.
The rule indicates that the new Dirac cones appear sporadically, which differ from the prediction of the previous study for periodic potentials \cite{park08}.
The sporadic appearance is due to the fact that the dispersion relation of graphene is linear only up to an energy cutoff.

Based on the results for the double-periodic potentials, we then predict that the new cones appear densely in the energy spectrum under quasiperiodic potentials.
The quasiperiodic potentials, a theoretical model of quasicrystal \cite{macia06}, are obtained as a limit of the double-periodic potentials.
Indeed, an experimental work recently reported quasiperiodic ripples in graphene grown by the chemical vapor deposition \cite{ni12}.
Quasiperiodic systems are widely considered to produce a fractal spectrum and vice versa \cite{macia06, suto89, guarneri94, tashima11}.
Graphene under a quasiperiodic potential of the form of the Fibonacci lattice indeed has been reported to have fractal structures in the electronic band gap and transport \cite{zhao11, sena10}.
Our study, however, suggests a \textit{non}-fractal appearance of the new Dirac cones in the energy spectrum.
The density of the new cones depend on the energy, again because of the energy cutoff of the linear dispersion.

We also extend the above results to materials with different values of the energy cutoff.
The generation of the new cones depend on the energy cutoff strongly.

\begin{figure}
\includegraphics[width=6.5cm, clip]{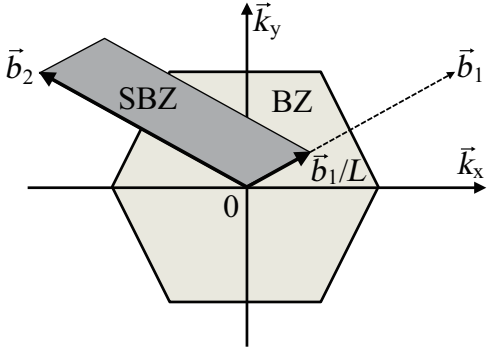}
\caption
{
The supercell Brillouin zone (SBZ) which we used (the dark gray area).
The light gray hexagon is the first Brillouin zone (BZ) of graphene without an external potential.
}
\label{fig:SBZ}
\end{figure}

\textit{Single-periodic potential}: Let us first review the prediction for single-periodic potentials \cite{park08, park08nat}.
We can approximate graphene under a superlattice potential $V(x,y)$ near the Fermi energy in the form \cite{park08, park08nat, sena10}:
\begin{equation}
		h \left[ V \right] = \hbar v_0 \left( - i \sigma_x \frac{\partial}{\partial x} -  i \sigma_y \frac{\partial}{\partial y}  \right) + I V (x,y) ,
\label{eq:dirac_form}
\end{equation}
where $v_0$ is the group velocity, $\sigma_x$ and $\sigma_y$ are the Pauli matrices, and $I$ is the identity matrix.
For simplicity, let us consider the potential varying only in the direction of $\vec{a}_1 = a / 2 (1, \sqrt{3})$ (where $a$ is the lattice constant): $V(\vec{r}) = V(\vec{r} + L \vec{a}_1)$, where $L$ is the period.
The Brillouin zone is then narrowed in the direction of $\vec{b}_1 = 2 \pi / (a \sqrt{3}) (\sqrt{3}, 1)$ in the reciprocal space as $(\vec{b}_1 / L) \times \vec{b}_2$ (Fig.~\ref{fig:SBZ}).
The reciprocal vectors of the potential are given by
\begin{equation}
	\vec{G}_0 (n) \equiv \frac{n}{L} \vec{b}_1 ,\quad n = 0, \pm 1, \pm 2, \dots \pm N.
\end{equation}
The new Dirac cones are predicted to appear around the boundary  $\vec{G}_0 (n) / 2$ of the supercell Brillouin zone (SBZ) \cite{park08}.
In our study we set the lattice constant $a$ of graphene, the hopping element $t_1$ of the tight-binding model on honeycomb lattice, and the Planck constant $\hbar$ all unity.
Then the group velocity $v_0$ is $(\sqrt{3} / 2) a t_1 = \sqrt{3} / 2$ and the energies of the new Dirac points are given by
\begin{equation}
	\frac{\hbar v_0 }{2} \frac{n}{L} \big| \vec{b}_1 \big| =  \frac{\pi}{L} n .
\end{equation}
Compared to the original Dirac cone ($n=0$), the new Dirac cones $( n \neq 0 )$ is skewed, its gradient in the $y$ direction being modified by the periodic potential \cite{park08}.

\begin{figure}
\includegraphics[width=6.5cm, clip]{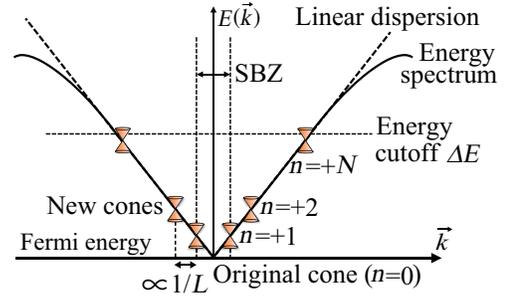}
\caption
{
A schematic illustration of the appearance of the new Dirac cones around the original Dirac point generated by a periodic potential.
The solid lines represent the dispersion of graphene, which deviates from the linear dispersion (the broken diagonal lines) beyond the energy cutoff $\Delta E$ (the broken horizontal line).
The broken vertical lines represent the SBZ.
The new cones are indexed as $n = 0, \pm 1, \dots, \pm N$, where $n=0$ corresponds to the original cone.
}
\label{fig:model}
\end{figure}

We numerically confirmed the above relation for the tight-binding model on a honeycomb lattice under a  sine function.
The result is  consistent with the positions reported in Ref.~\cite{park08}.
The indices $n = 0, \pm 1, \dots, \pm N$ are consecutive up to an energy cutoff $\Delta E \simeq 0.62$, beyond which the dispersion of graphene cannot be approximated to be linear and the new cones are not generated anymore (see Fig.~\ref{fig:model}).
The maximum index is therefore $N = \lfloor ( L \, \Delta E ) / \pi \rfloor$, where $\lfloor \cdot \rfloor$ is the integer portion of a number.

\textit{Double-periodic potential}: We next see what happens when we apply a double-periodic potential to graphene.
A double-periodic function is defined as follows.
For simplicity again, let us consider the functions $V_1$ and $V_2$ which are periodic along the $\vec{a}_1$ axis,
\begin{equation}
V_{i} (\vec{r} + L_i \vec{a}_1) = V_{i} (\vec{r})
\label{eq:potentials}
\end{equation}
for $i=1,2$, where the integers $L_1$ and $L_2$ are the periods of the two functions and {\it coprime}.
We refer to the sum $V = V_1 + V_2$ as a double-periodic function with period $L = L_1 \times L_2$.
The double-periodic function is obviously a periodic function.
One would therefore expect to understand the generation of new cones in terms of the single-periodic superlattice potential and would predict that the new Dirac points appear in the energy spectrum at
\begin{equation}
	\frac{\pi}{L} n_\textrm{d} , \quad n_\textrm{d} = 0, \pm 1, \pm 2, \dots , N_\textrm{d}
\label{eq:position_L} 
\end{equation}
consecutively up to the limit $N_\textrm{d} = \lfloor (L_1 L_2 \Delta E ) / \pi \rfloor$.

However, our numerical results for the tight-binding model under double-periodic potentials indicate otherwise; the new cones do not appear consecutively, only sporadically.
We hereafter introduce an additional rule governing the sporadic appearance.

The reciprocal vector of each function $V_i$ are given by 
\begin{equation}
	\vec{G}_{i} (n_i) \equiv \frac{n_i}{L_i} \vec{b}_1, \quad n_i = 0, \pm1, \pm 2, \dots, \pm N_i,
\end{equation}
where $N_i = \lfloor (L_i \Delta E) / \pi \rfloor$ ($i = 1, 2$).
A similar argument to Ref.~\cite{park08} predicts the energies of the new Dirac points as
\begin{equation}
	\frac{\hbar v_0}{2} \left( \frac{n_1}{L_1} \big| \vec{b}_1 \big| + \frac{n_2}{L_2} \big| \vec{b}_1 \big| \right) = \left( \frac{\pi}{L_1} n_1 + \frac{\pi}{L_2} n_2 \right) 
\label{eq:position_L12}
\end{equation}
for the linearized Hamiltonian \eqref{eq:dirac_form}.
Equating Eqs.~\eqref{eq:position_L} and \eqref{eq:position_L12}, we obtain the Diophantine equation:
\begin{equation}
n_\textrm{d} = L_2 n_1 + L_1 n_2 .
\label{eq:cone_indices}
\end{equation}
The point is the following. 
If the indices $n_1$ and $n_2$ took all integer values as is for the Dirac Hamiltonian \eqref{eq:dirac_form}, the Diophantine equation \eqref{eq:cone_indices} could produce any integers for $n_\textrm{d}$.
For graphene, however, each of $n_1$ and $n_2$ is consecutive only up to its respective limit $N_1$ or $N_2$, and therefore the index $n_\textrm{d}$ is \textit{not} always consecutive.

Let us exemplify the above by numerically diagonalizing the tight-binding model on a honeycomb lattice under a double-periodic potential.
We defined the double-periodic function $V$ as the sum of two sine functions with the periods $(L_1, L_2) = (13,8)$, where the unity means the normalized lattice constant.
Therefore, the total period of the double-periodic potential is given by $L = 104$. 
We set the amplitudes of the potentials as $v_1 = 0.7$ and $v_2 = 0.25$.

First, the numerical results for graphene under only one of the sine potentials $V_1$ and $V_2$ confirm that the indices $n_1$ and $n_2$ are limited to $n_1 = 0, \pm 1, \pm 2$ and $n_2 = 0, \pm 1$, respectively, because of the restrictions $\big| n_i \big| \le \lfloor (L_i \Delta E ) / \pi \rfloor$ ($i=1,2$) with $\Delta E \simeq 0.62$. 
We show in Table~\ref{tab:table1} our prediction of all possible values of the index $n_\textrm{d}$ according to the Diophantine equation~\eqref{eq:cone_indices}; we focus on the positive energy range because the energy spectrum is symmetric about the Fermi energy.
\begin{table}
\caption{
All possible values of the index $n_\textrm{d}$ according to Eq.~\eqref{eq:cone_indices}, compared with the numerical data of the new cones in graphene under a double-periodic potential  with $L_1 = 13$ and $L_2 = 8$. 
The first two columns show all combinations of the indices $n_1 = 0, \pm1, \pm2$ and $n_2 = 0, \pm1$ for each sine potential and the corresponding index $n_d$ of the new cones.
The new cones are predicted to appear around the positions $\left(\pi / 104 \right) \times n_\textrm{d}$ in the energy spectrum, which the numerical data confirm. 
We used the units $t_1 = a = \hbar = 1$.
}
\begin{ruledtabular}
\begin{tabular}{lcll}
$(n_1 , n_2)$ & $n_\textrm{d}$ in Eq.~\eqref{eq:cone_indices} & $(\pi / 104) \times n_\textrm{d}$ & numerical data\\
\hline
$(+2 , -1)$ & $+3$ & $\simeq 0.09062$ & 0.0775\\
$(-1 , +1)$ & $+5$ & $\simeq 0.1510$ & 0.149 \\
$(+1 , 0)$ & $+8$ & $\simeq 0.2417$ & 0.247 \\
$(0 , +1)$  & $+13$ & $\simeq 0.3927$ & 0.356 \\
$(+2 , 0)$  & $+16$ & $\simeq 0.4833$ & 0.468 \\
$(+1 , +1)$ & $+21$ & $\simeq 0.6344$ & not observed \\
$(+2, +1)$   & $+29$ & $\simeq 0.8760$ & not observed \\
\end{tabular}
\end{ruledtabular}
\label{tab:table1}
\end{table}

\begin{figure}
\includegraphics[width=6.5cm, clip]{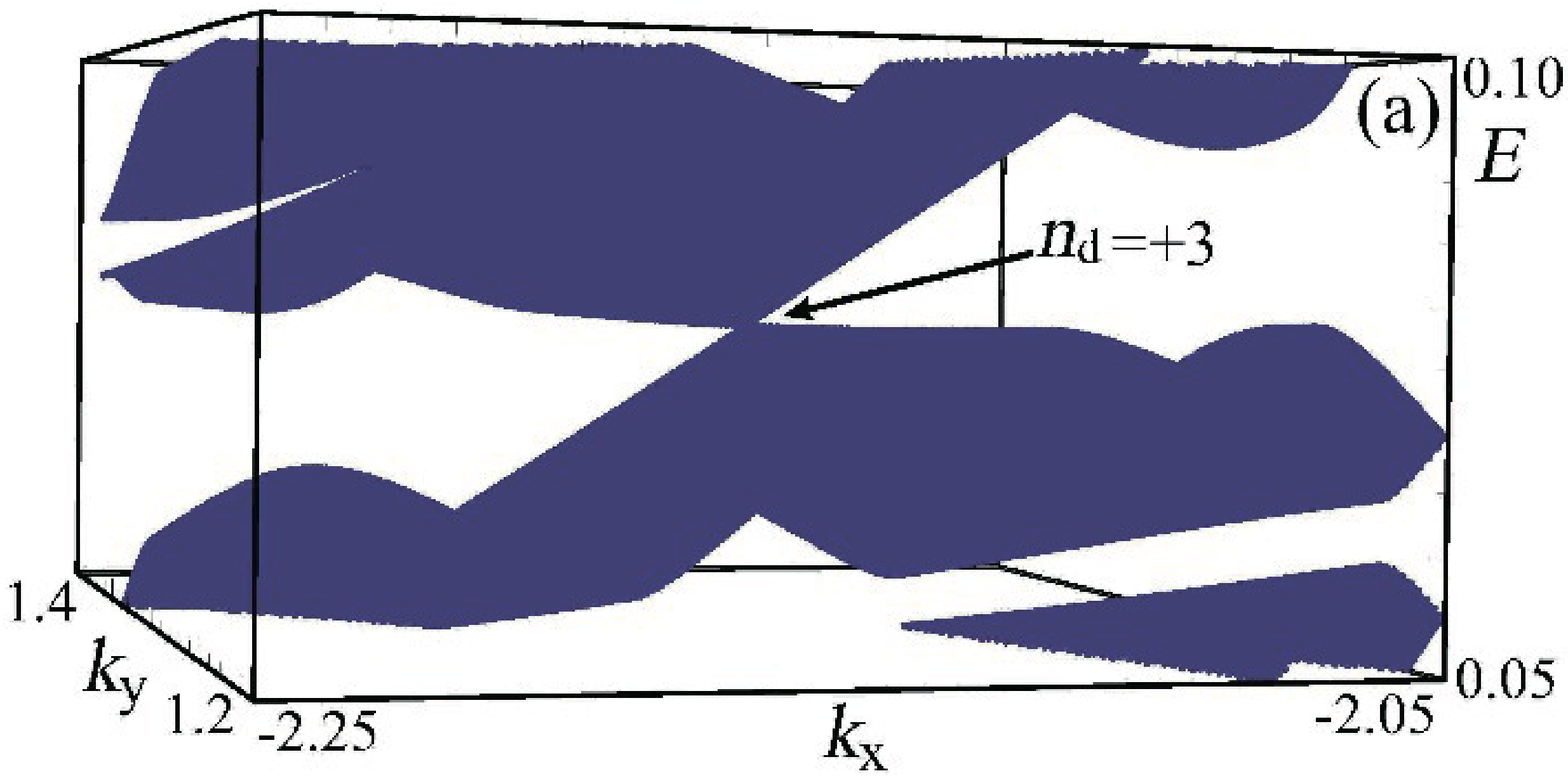}
\includegraphics[width=6.5cm, clip]{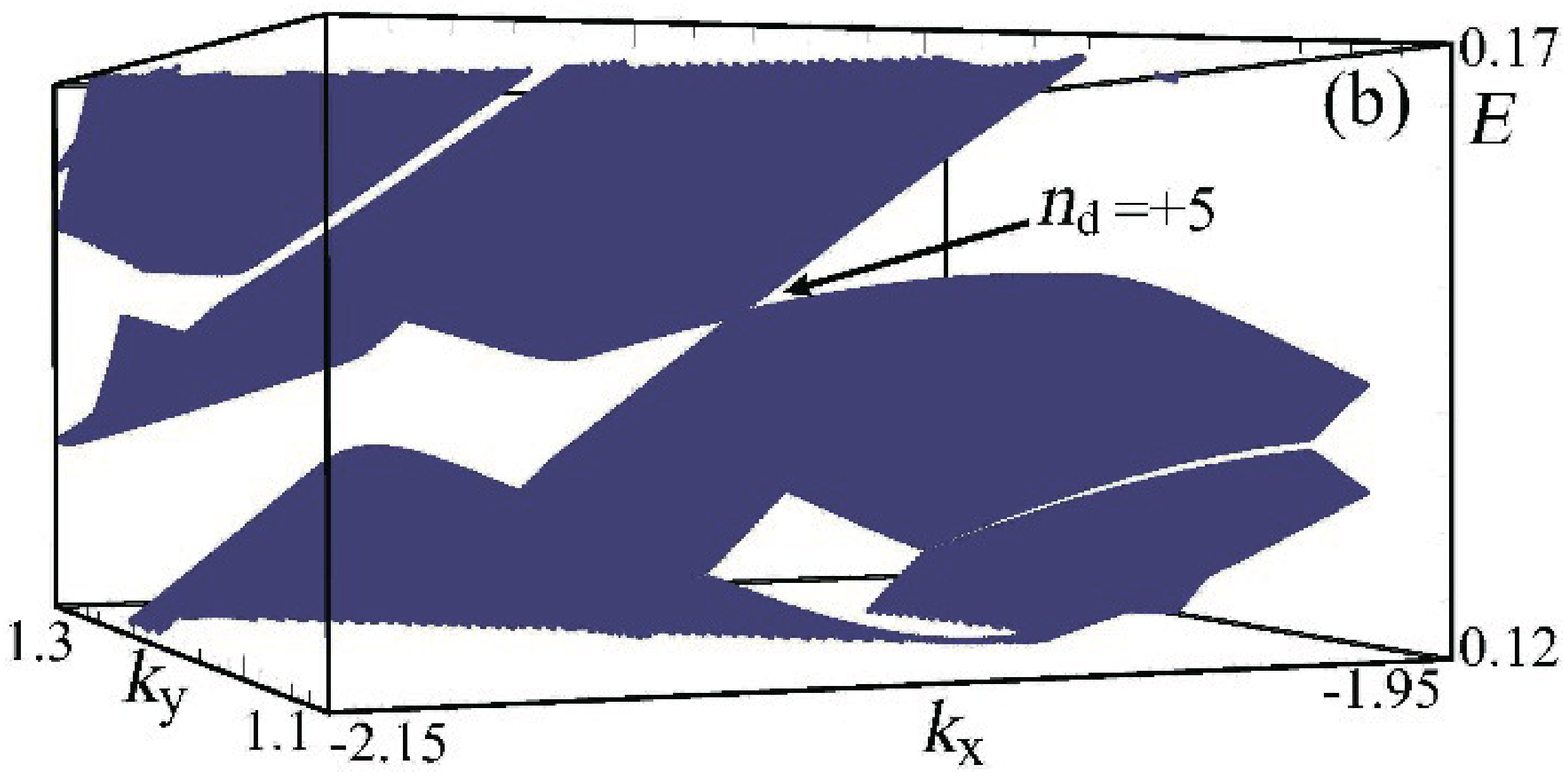}
\caption
{
The energy spectrum of graphene (the tight-binding model on a honeycomb lattice) under a double-periodic potential with $L_1 = 13$, $L_2 = 8$, $v_1 = 0.7$, and $v_2 = 0.25$.
The energy ranges of \textbf{(a)} $0.05 \le E \le 0.1$  and \textbf{(b)} $0.12 \le E \le 0.17$ are illustrated.
The original Dirac cone is located at the K point $(k_x , k_ y) \simeq (-2.17, 1.27)$. 
The first and second new cones are clearly observed at \textbf{(a)} $E \simeq 0.0775$ and \textbf{(b)} $E \simeq 0.149$, which we indicated with arrows.
}
\label{fig:fig1}
\end{figure}

Our numerical results for the double-periodic potential (Fig.~\ref{fig:fig1}) agree well with the predictions in Table~\ref{tab:table1}.
Their deviations are probably due to the deviation of the energy spectrum of pure graphene from the linear dispersion.
The last two cones $n_\textrm{d} = + 21$ and $+ 29$ in Table \ref{tab:table1} did not appear because their energies are greater than the cutoff $\Delta E \simeq 0.62$.
To summarize, the indices of the generation rule \eqref{eq:cone_indices} must satisfy the following restrictions: $|n_1| \le N_1$, $|n_2| \le N_2$, and $|n_\textrm{d}| \le N_\textrm{d}$.

The generation rule \eqref{eq:cone_indices} did not appear in the previous study \cite{park08,park08nat} presumably because of  the shape of the external potential. 
Let us consider the case in which $L_1$ and $L_2$ are not coprime, e.g. $L_1 = 16$ and $L_2 = 8$.
Because the total period is $L = 16$, Eqs.~\eqref{eq:position_L} and \eqref{eq:position_L12} gives $n_\textrm{d} = n_1 + 2 n_2$.
The cone index $n_\textrm{d}$ would take consecutive integers up to $N_\textrm{d} = 3$ with $| n_1 | \le 3$ and $| n_2 | \le 1$. 
This illustrates the special aspect of double-periodic potentials with coprime integers $L_1$ and $L_2$.

\begin{figure}
\includegraphics[width=6.5cm, clip]{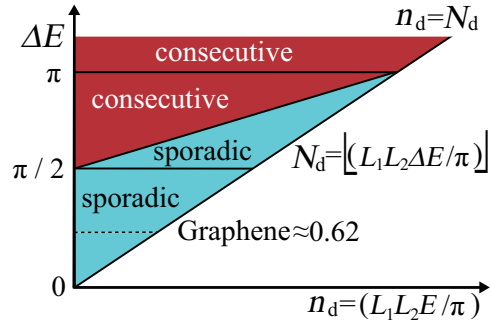}
\caption
{
A schematic of the appearance of the new Dirac points.
The dark red area is filled up by the new points.
In the light blue area the points appear sporadically.
The broken line represents the energy cutoff of graphene (the tight-binding model on a honeycomb lattice).
}
\label{fig:pa}
\end{figure}

In order to generalize the argument and take account of various materials with Dirac cones, let us make the energy cutoff $\Delta E$ a free parameter.
We have three cases of the energy cutoff $\Delta E$ regarding the appearance of the new cones (Fig.~\ref{fig:pa}).
In the first case $\Delta E \ge \pi $, all new cones appear consecutively up to $N_\textrm{d}$, which coincides with the case of a single sine function.
The second case is for the cutoff $\pi / 2 \le \Delta E < \pi$, where the new cones fill up the energy range $|E| \le 2 \Delta E - \pi$, or $| n_\textrm{d} | \leq ((2 \Delta E)/\pi - 1) L$.
In the energy range $2 \Delta E - \pi < |E| \le \Delta E$, the new cones appear only sporadically.
In the last case $\Delta E < \pi /2$, the new cones appear sporadically in all energy range $|E| \le \Delta E$. 
Graphene corresponds to the third case.

We can explain these three cases with the finite simple continued-fraction expansion \cite{jones80} of the rational number $r = L_1 / L_2 $.
The Euclidian algorithm \cite{cohen93} casts any rational number into two types of the continued-fraction expansion $r = [b_0, b_1, \dots, b_\nu] $ and $[b_0, b_1, \dots, (b_\nu - 1), 1] $, where $\{ b_\nu \}$ are positive integers.

We rewrite the generation rule \eqref{eq:cone_indices} by applying the expansion as well as the restrictions of the indices, $|n_1| \le N_1 = \lfloor ( L_1 \Delta E) / \pi \rfloor$ and $|n_2| \le N_2 = \lfloor ( L_2 \Delta E ) / \pi \rfloor$, obtaining two inequalities in terms of the coefficients of the continued-fraction expansion:
\begin{equation}
	\left| \frac{A_{\nu-1}}{A_\nu} n_\textrm{d} -m \right| \le \frac{N_1}{A_\nu}, \quad \left| \frac{B_{\nu-1}}{B_\nu} n_\textrm{d} -m \right| \le \frac{N_2}{B_\nu},
\label{eq:inequalities}
\end{equation}
where $A_\nu = L_1$, $B_\nu = L_2$, $A_{\nu - 1}$ and $B_{\nu - 1}$ are integers satisfying $A_{\nu - 1} / B_{\nu - 1} = [b_0, b_1 , \dots, b_{\nu - 1}]$, and $m$ is an arbitrary integer.
The two types of the expansion yield the same inequalities \eqref{eq:inequalities}.
Only the cone index $n_\textrm{d}$ which satisfies both inequalities gives a new Dirac cone.
The inequalities \eqref{eq:inequalities} tell us that the new cones appear consecutively in the range $|E| \le 2 \Delta E - \pi$ and sporadically in the range $2 \Delta E - \pi < |E| \le \Delta E$.

\textit{Quasiperiodic potential}: We are now in a position to study graphene under quasiperiodic potentials.
The sum $V = V_1 + V_2$ becomes quasiperiodic when the ratio $r = L_1 / L_2$ is irrational \cite{macia06}.
We can approximate the quasiperiodic functions by the double-periodic functions as follows.
Any irrational number $r_\infty$ can be represented by an infinite simple continued fraction \cite{jones80}
\begin{equation}
r_\infty = [b_0, b_1, b_2, \dots, b_\nu, \dots] , 
\end{equation}
where $\{ b_\nu \}$ are positive integers.
A rational number $r_\nu = [b_0 , b_1 , \dots, b_\nu] = A_\nu / B_\nu$  converges to the irrational number $r_\infty$ in the limit $\nu \to \infty$ .
Then the double-periodic potential with $L_1 = A_\nu$ and $L_2 = B_\nu$ approximates the quasiperiodic potential. 
For example, the golden ratio is approximated by the series of rational numbers, $A_1 / B_1 = 2/1$, $A_2 / B_2 = 3/2$, $A_3 / B_3 = 5/3$, $A_4 / B_4 = 8/5$, $A_5 / B_5 = 13/8$, and so on.

\begin{figure}
\includegraphics[width=7.cm, clip]{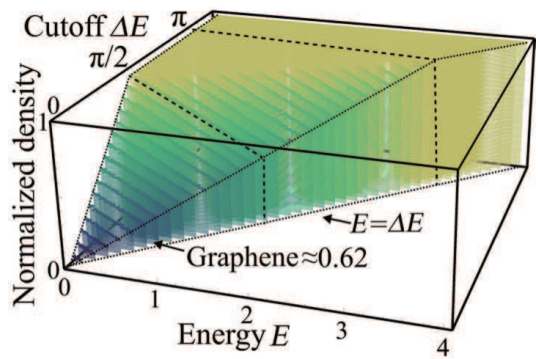}
\caption
{
The normalized density for quasiperiodic potentials, with an example of $L_1 / L_2 = 233 /144$ as an approximation of the golden ratio.
The two-dimensional plane represents the energy $E$ and the energy cutoff $\Delta E$.
The dotted lines represent the general results derived from Eq.~\eqref{eq:inequalities}.
The broken lines represent the special cases $\Delta E = \pi /2$ and $\pi$.
}
\label{fig:density}
\end{figure}

The existence of the solutions is basically the same as shown in Fig.~\ref{fig:pa}, except that the new cones appear densely in the energy range $|E| \le \Delta E$ because ($L_1$, $L_2$) $=$($A_\nu$, $B_\nu$) $\to \infty$ in the quasiperiodic limit $\nu \to \infty$.
The three cases above are now distinguished in terms of the density $\rho_{\textrm{Dirac}}$ of the Dirac cones.
Let us normalize the density of the new cones by the density in the case of a single sine function, namely $L / \pi$.
We show in Fig.~\ref{fig:density} the normalized density of the new cones for quasiperiodic potentials, with an example of $L_1 = 233$ and $L_2 = 144$, which emulate the golden ratio.
In the first case $\Delta E \ge \pi$, the normalized density is unity $\rho_{\textrm{Dirac}} = 1$ in the whole range $|E| \le \Delta E$.
In the second case $\pi / 2 \le \Delta E < \pi$, the inequalities \eqref{eq:inequalities} tell us that $\rho_{\textrm{Dirac}} = 1$ for $|E| \le 2 \Delta E - \pi$ but $ \rho_{\textrm{Dirac}} = ( 2 \Delta E - E ) / \pi < 1$ for $2 \Delta E - \pi < |E| \le \Delta E$.
In the third case $\Delta E < \pi / 2$, the density is always less than unity, $ \rho_{\textrm{Dirac}} = ( 2 \Delta E - E ) / \pi $.
Graphene falls into the third case.

We have carried out a multifractal analysis for the sporadic series of the new cones in the second and the third cases.
The multifractal spectrum seems to converge to one point with the fractal dimension one as the order of the expansion $\nu$ increases.
This fact indicates that the appearance of the new cones are {\it not} fractal in the quasiperiodic limit $\nu \to \infty$ in the second and the third cases.
A Fourier analysis of the interval of the new cones suggests the same.
We therefore conclude that the new cones appear almost regularly, although the density of the cones are less than in the first case.

\textit{Summary}: 
We found a new generation rule of the Dirac cones for graphene under double-periodic and quasiperiodic potentials on the basis of the Diophantine equation.
The generation of the new Dirac points is classified into three cases in terms of the density, depending on the energy cutoff.
We also showed that the appearance of the new Dirac points under a quasiperiodic potential is not fractal.
These results will be important in understanding the fundamental natures of graphene and other Dirac electron systems under external potentials.

\begin{acknowledgments}
We thank Professor J. Goryo for useful discussions.
We are indebted to Dr. K. Imura for fruitful information on this work.
This work is supported by Grant-Aid for scientific Research No.\ 17340115 from the Ministry of Education, Culture, Sports, Science and Technology.
One of the authors, M.T., is supported by Global Center of Excellence for Physical Sciences Frontier, The University of Tokyo .
\end{acknowledgments}



\end{document}